\documentclass{aa}  
\usepackage{epsfig}
\usepackage{graphicx}
\usepackage{xcolor}
\usepackage{txfonts}
\usepackage{amssymb}
\usepackage{amsfonts}
\usepackage{natbib}
\usepackage{hyperref}
\hypersetup{
  colorlinks=true, 
  urlcolor=blue, 
  linkcolor=blue, 
  citecolor=blue 
}

\bibpunct{(}{)}{;}{a}{}{,}

\def\etal{{et\,al.}\ }

\newcounter{Rco}

\newcommand{\logg}{\mbox{$\log g$}\xspace}

\newcommand{\Teff}{\mbox{$T_\mathrm{eff}$}\xspace}

\newcommand{\Lsol}{\ensuremath{L_\odot}}
\newcommand{\Msol}{\ensuremath{M_\odot}}
\newcommand{\Rsol}{\ensuremath{R_\odot}}
\newcommand{\Mdot}{\ensuremath{\dot{M}}}

\def\civ{\ion{C}{IV}}

\def\oiii{\ion{O}{III}}

\newcommand{\Gaia}{{\it Gaia}}
\newcommand{\GALEX}{{\it GALEX}}
\newcommand{\WISE}{{\it WISE}}

\begin{document}

%

\title{Spectroscopic survey of faint planetary-nebula nuclei \newline  III\null. A [WC] central star and two new PG1159 nuclei\thanks{Based on observations obtained with the Hobby-Eberly Telescope (HET), which is a joint project of the University of Texas at Austin, the Pennsylvania State University, Ludwig-Maximillians-Universit\"at M\"unchen, and Georg-August Universit\"at G\"ottingen. The HET is named in honor of its principal benefactors, William P. Hobby and Robert E. Eberly.} }

\author{
Klaus Werner\inst{1} 
\and Helge Todt \inst{2}
\and Howard E. Bond\inst{3,}\inst{4}
\and Gregory R. Zeimann\inst{5}
}

\institute{Institut f\"ur Astronomie und Astrophysik, Kepler Center for
  Astro and Particle Physics, Eberhard Karls Universit\"at, Sand~1, 72076
  T\"ubingen, Germany\\ \email{werner@astro.uni-tuebingen.de} 
\and
Institut f\"ur Physik und Astronomie, Universit\"at Potsdam, Karl-Liebknecht-Stra\ss e 24/25, 14476 Potsdam, Germany
\and
Department of Astronomy \& Astrophysics, Pennsylvania State University, University Park, PA 16802, USA
\and
Space Telescope Science Institute, 3700 San Martin Dr., Baltimore, MD 21218, USA
\and
Hobby-Eberly Telescope, University of Texas at Austin, Austin, TX 78712, USA
}

\date{Received 15 January 2024 / Accepted 29 February 2024}

\authorrunning{K. Werner \etal}
\titlerunning{New hydrogen-deficient central stars}

\abstract{
We present spectroscopy of three hydrogen-deficient central stars of faint planetary nebulae, with effective temperatures (\Teff) in excess of $100\,000$\,K. The nucleus of RaMul~2 is a Population~II Wolf-Rayet star of spectral type [WC], and the central stars of Abell~25 and StDr~138 are two new members of the PG1159 class. Our spectral analyses reveal that their atmospheres have a similar chemical composition. They are dominated by helium and carbon, which was probably caused by a late helium-shell flash. Coincidentally, the three stars have similar masses of about $M=0.53$\,\Msol\ and, hence, form a post-AGB evolutionary sequence of an initially early-K type main-sequence star with $M=0.8$\,\Msol. The central stars cover the period during which the luminosity fades from about 3000 to 250\,\Lsol\ and the radius shrinks from about 0.15 to 0.03\,\Rsol. The concurrent increase of the surface gravity during this interval from \logg = 5.8 to 7.2 causes the shutdown of the stellar wind from an initial mass-loss rate of $\log \dot{M}/(\Msol\,{\rm yr}^{-1}) = -6.4$, as measured for the [WC] star. Along the contraction phase, we observe an increase of \Teff from 112\,000\,K, marked by the [WC] star, to the maximum value of 140\,000\,K and a subsequent cooling to 130\,000\,K, marked by the two PG1159 stars. 
}


\maketitle
%

\section{Introduction} \label{sect:intro}

This is the third in a series of papers presenting results from a spectroscopic survey of central stars of faint Galactic planetary nebulae (PNe). It is being carried out with the second-generation Low-Resolution Spectrograph (LRS2; \citealt{Chonis2016}) of the 10-m Hobby-Eberly Telescope (HET; \citealt{Ramsey1998,Hill2021}), located at McDonald Observatory in west Texas, USA\null. An overview of the survey, a description of the instrumentation and data-reduction procedures, target selection, and some initial results were presented in our first paper \citep[][hereafter Paper~I]{Bond2023a}. Paper~II in this series \citep{Bond2023b} discussed the central star of the ``PN mimic'' Fr~2-30. In this third paper we present spectra of three little-studied objects that have extremely hot, hydrogen-deficient central stars. About 50 central stars have been observed to date. Future papers will discuss several more individual objects of special interest, and another publication will present results on a group of nuclei with fairly normal hydrogen-rich spectra.

The remainder of this paper is organized as follows. In Sect.\,\ref{sect:targets} we introduce our program stars and their planetary nebulae, and we describe our spectroscopic observations in Sect.\,\ref{sect:observations}. In Sect.\,\ref{sect:analysis} we 
present our spectral analyses and their results. We summarize and discuss our findings in Sect.\,\ref{sect:summary}.

\section{The targets} \label{sect:targets}

Table~\ref{tab:DR3data} lists celestial and Galactic coordinates, parallaxes, and magnitudes and colours for our three central stars, all taken from \Gaia\/ Data Release~3\footnote{\url{https://vizier.cds.unistra.fr/viz-bin/VizieR-3?-source=I/355/gaiadr3}} (DR3; \citealt{Gaia2016, Gaia2023}). The following subsections give brief details of the discoveries of these faint PNe and their nuclei, and some of the nebular properties. Further information about the objects is contained in the online Hong-Kong/AAO/Strasbourg/H$\alpha$ Planetary Nebulae (HASH) database\footnote{\url{http://hashpn.space/}} \citep{Parker2016}.

\begin{table*}
\centering
\caption{\emph{Gaia}/DR3 data for the central stars of Abell 25, StDr 138, and RaMul 2.}
\label{tab:DR3data}
\begin{tabular}{lccc} 
	\hline\hline
Parameter                     & Abell 25          & StDr 138         & RaMul 2 \\
	\hline
\Gaia\/ DR3 ID                & 3070613063558890240 & 238382646416730368 & 1821559921653783552 \\ 
RA (J2000)                    & 08 06 46.501      & 03 37 55.777     & 19 49 53.705\\
Dec (J2000)                   & $-$02 52 35.18    & $+43$ 44 15.74  & $+$18 40 14.82\\
$l$ [deg]                     &  224.39           & 152.05           & 056.18\\
$b$  [deg]                    &  $+15.31$         & $-09.52$         & $-$03.83\\
Parallax [mas]                & $0.386\pm0.163$   & $1.022\pm0.150$  & $0.108\pm0.035$\\
$\mu_\alpha$ [mas\,yr$^{-1}$] & $-1.439\pm0.160$ & $-4.839\pm0.154$ & $-1.792\pm0.033$\\
$\mu_\delta$ [mas\,yr$^{-1}$] & $+3.532\pm0.099$ & $-0.481\pm0.141$ & $-3.681\pm0.032$\\
$G$ [mag]                     &  18.38            & 17.96            & 15.60  \\
$G_{\rm BP}-G_{\rm RP}$ [mag] & $-0.54$           & $-0.13$	         & 0.31   \\
 	\hline
\end{tabular}
\end{table*}

\subsection{Abell~25 (K 1-13)}

This low-surface-brightness nebula (PN G224.3+15.3) was discovered six decades ago by \cite{1963BAICz..14...70K} in his inspection of prints from the Palomar Observatory Sky Survey (POSS), and is designated K\,1-13. It also appears as entry number 25 in the classical list of ancient PNe found on the POSS by \cite{Abell1966}. We will use the designation Abell~25 hereafter, as this name has been used more commonly in the PN literature. 

The nebula has a barrel-shaped bipolar morphology with relatively bright condensations on either side of the central star, separated by a width of about $150''$. At the distance given by the \Gaia\/ parallax, this corresponds to a physical width of about 1.9~pc. Several narrow-band images of Abell~25 obtained by advanced amateurs are available, one of the best being a deep frame by Eduardo Rigoldi.\footnote{\url{https://www.astrobin.com/nk55mw/}} The blue 18th-mag central star was pointed out by Kohoutek and Abell; much more recently, it was listed as a white-dwarf (WD) candidate by \cite{2019MNRAS.482.4570G,2021MNRAS.508.3877G} with the name 
WD~J080646.50$-$025235.18. It is also contained in catalogues of hot subluminous stars assembled from {\it Gaia\/} data by \citet{Geier2019} and \citet{Culpan2022}. 


\subsection{StDr~138 \label{subsec:targetStDr138} }

This object (PN~G152.0$-$09.5) came to our attention in 2021 April because of an image posted to an online amateur astrophotography group, showing a newly discovered large and very low-surface-brightness nebula. Our inspection of sky-survey images revealed a 17th-mag blue star near its center, making it the probable nucleus. A redder and 0.9~mag brighter star lies only $4\farcs9$ away, but its \Gaia\/ parallax and proper motion show it to be an unrelated background star. A subsequent literature search found that our central star had been listed as a WD candidate in searches of the \Gaia\/ catalogue by \cite{2019MNRAS.482.4570G,2021MNRAS.508.3877G}, with the designation WD~J033755.78$+$434415.74. The high temperature of the star is confirmed by its presence in the \GALEX\/ point-source catalogue\footnote{\url{https://galex.stsci.edu/GR6/}} as \GALEX\/ J033755.8+434415, with FUV and NUV magnitudes of 17.38 and 17.98, respectively.

The discovery of StDr~138 was formally announced in a paper by \citet{2022A&A...666A.152L}, which describes extensive searches for faint PNe by amateur astronomers, primarily in France.\footnote{See \url{http://planetarynebulae.net}} The object was discovered by Xavier Strottner and Marcel Drechsler, and has an angular size of $9.5 \times 12$ arcmin. The corresponding physical dimensions are about $2.7\times3.4$~pc---suggesting that StDr~138 may be older than Abell~25. A few deep images are available online, of which the most extraordinary is one obtained by Nicolas Outters.\footnote{\url{https://www.astrobin.com/1dzqtd/}} It shows an elongated PN with two relatively bright regions on both sides of the central star (somewhat reminiscent of Abell~25), enclosed in a large and faint elliptical shell.
Although
described as an oval PN in HASH, StDr~138 could be a bipolar nebula highly
inclined to the line of sight. A kinematic study would help clarify this.



\subsection{RaMul~2 \label{subsec:targetRaMul2 }}

As described in \citet{AckerLeDu2015}, RaMul~2 (PN G056.1$-$03.8) was discovered independently by amateurs Thierry Raffaelli and later by Lionel Mulato, the latter by searching mid-infrared images from the \WISE\/ spacecraft.\footnote{Mulato tells the story of his discovery at \url{https://tinyurl.com/mn3hb4at}} A spectrum obtained in 2015 by amateur Christian Buil (presented by \citealt{Acker2015}, and also available at \url{planetarynebulae.net}) showed broad emission features of \civ\ in addition to [\oiii] nebular lines, indicating that the object is a PN with a Wolf-Rayet (WR) central star. \citet{Acker2015} assigned it a spectral type of [WO4]. Our attention was drawn to the object by a conference paper by \citet{LeDu2018}, which presented amateur discoveries of several new PNe. In the publications cited here, as well as in the SIMBAD database,\footnote{\url{http://simbad.u-strasbg.fr/simbad/sim-fid}} the object is designated ``Mul~5.'' However, in the HASH catalogue, ``Mul~5'' refers to a completely different object, and the designation RaMul~2 is given to PN~G056.1$-$03.8, in order to recognize the independent discovery by Raffaelli. We will use this name here.

Sky-survey images show the PN to be a thin elliptical ring with dimensions of $17''\times26''$ (roughly $0.5\times0.8$~pc, making the object likely the youngest of the three targets in this paper). The PN hosts a prominent blue central star. RaMul~2 lies at a low Galactic latitude, raising the possibility that its nucleus could be a massive Population~I WR star, rather than a true PN central star. In fact, as noted by Mulato, the object lies within about half a degree of a well-known WR star, WR~128. However, the faintness of the RaMul~2 central star, and its \Gaia\/ parallax and proper motion, are inconsistent with the high luminosity of a massive WR star.

\section{Observations and Data Reduction}\label{sect:observations}

As noted in the Introduction, our spectroscopic observations were made with the LRS2 spectrograph on the HET at McDonald Observatory. Paper~I gives details of the instrumentation. We note here that LRS2 is composed of two integral-field-unit arms, blue (LRS2-B) and red (LRS2-R), but all of our observations were made with the targets placed in the LRS2-B arm. It employs a dichroic beamsplitter to send light simultaneously into two spectrograph units: the ``UV'' channel (covering 3640--4645~\AA\ at resolving power 1910), and the ``Orange'' channel (covering 4635--6950~\AA\ at resolving power 1140). The data were initially processed using \texttt{Panacea},{\footnote{\url{https://github.com/grzeimann/Panacea}} which performs bias and flat-field correction, fiber extraction, and wavelength calibration. An absolute-flux calibration comes from default response curves and measures of the mirror illumination, as well as the exposure throughput from guider images.  We then applied \texttt{LRS2Multi}\footnote{\url{https://github.com/grzeimann/LRS2Multi}} to the un-sky-subtracted, flux-calibrated fiber spectra, to perform background and sky subtraction in an annular aperture, source extraction using a 2$\arcsec$ radius aperture, and combination of multiple exposures if applicable, similar to the description in Paper~II\null. The final spectra were resampled to a common linear grid with a 0.7~\AA\ spacing, and then normalized to a flat continuum for further analysis. The final spectra are displayed below in Figs.\ref{fig:a25}--\ref{fig:ramul2}. An observation log for our LRS2-B exposures is presented in Table~\ref{tab:exposures}.

\begin{table}
\centering
\caption{HET LRS2-B observations.}
\label{tab:exposures}
\begin{tabular}{lcc} 
        \hline\hline 
Name &          Date    &       Exposure \\
     &     [YYYY-MM-DD] & [s]            \\
        \hline
Abell 25     &  2019-11-01   & $2\times450$  \\
             &  2021-02-07   & $2\times900$  \\
             &  2021-11-30   & $2\times1000$ \\
StDr 138     &  2021-11-02   & $2\times600$  \\
RaMul 2      &  2023-03-31   & 180           \\
        \hline
\end{tabular}
\end{table}

\begin{figure*}
 \centering  \includegraphics[width=0.9\textwidth]{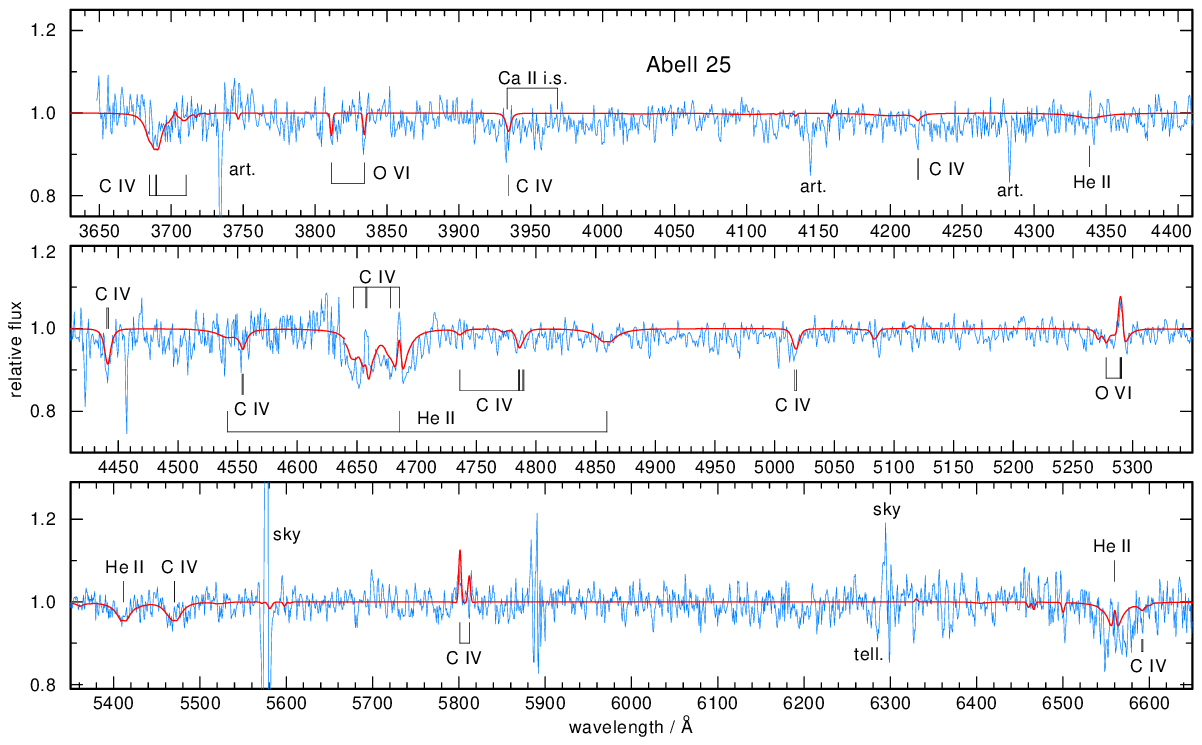}
 \caption{HET/LSR2 spectrum (blue line) of the PG1159 central star of Abell~25. Overplotted (red line) is our best-fit model with \Teff = 140\,000\,K, \logg = 6.8, and abundances of He = 0.46, C = 0.46, and O = 0.08.}
\label{fig:a25}
\end{figure*}

\begin{figure*}
 \centering  \includegraphics[width=0.9\textwidth]{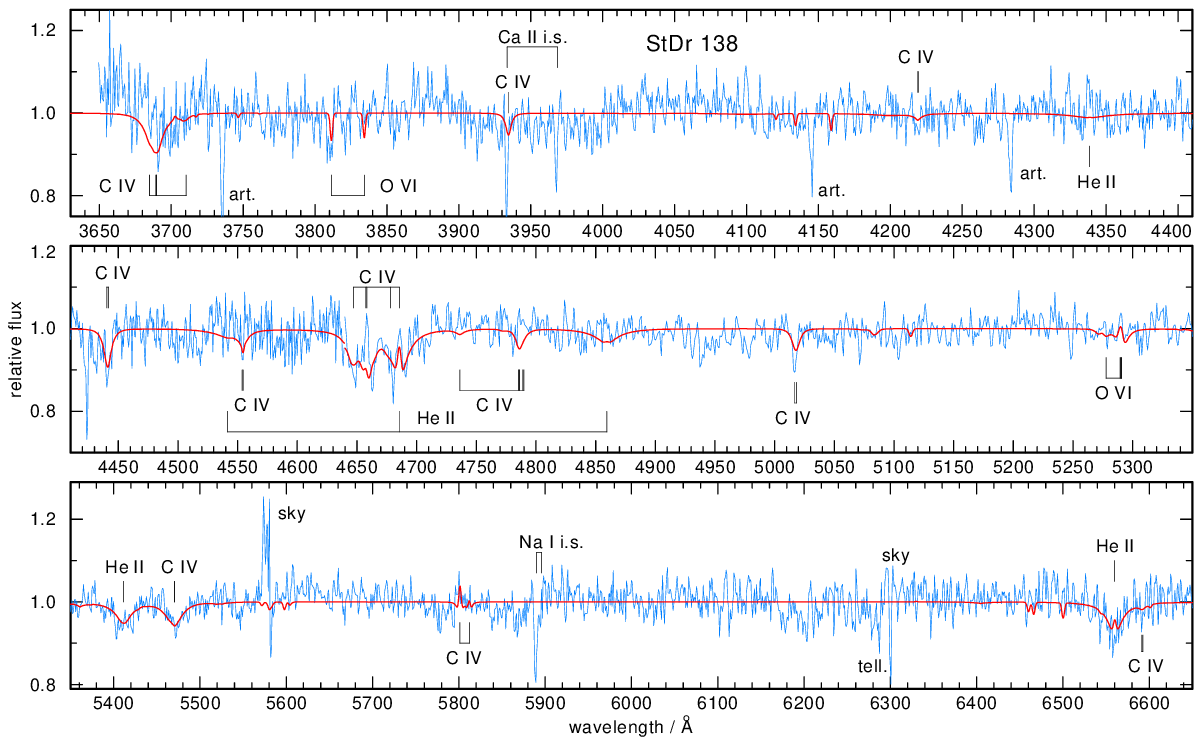}
 \caption{HET/LSR2 spectrum (blue line) of the PG1159 central star of StDr~138. Overplotted (red line) is a model with \Teff = 130\,000\,K, \logg=7.2, and abundances of He = 0.46, C = 0.46, and O = 0.08.}
\label{fig:stdr138}
\end{figure*}

\begin{figure*}
 \centering  \includegraphics[width=0.9\textwidth]{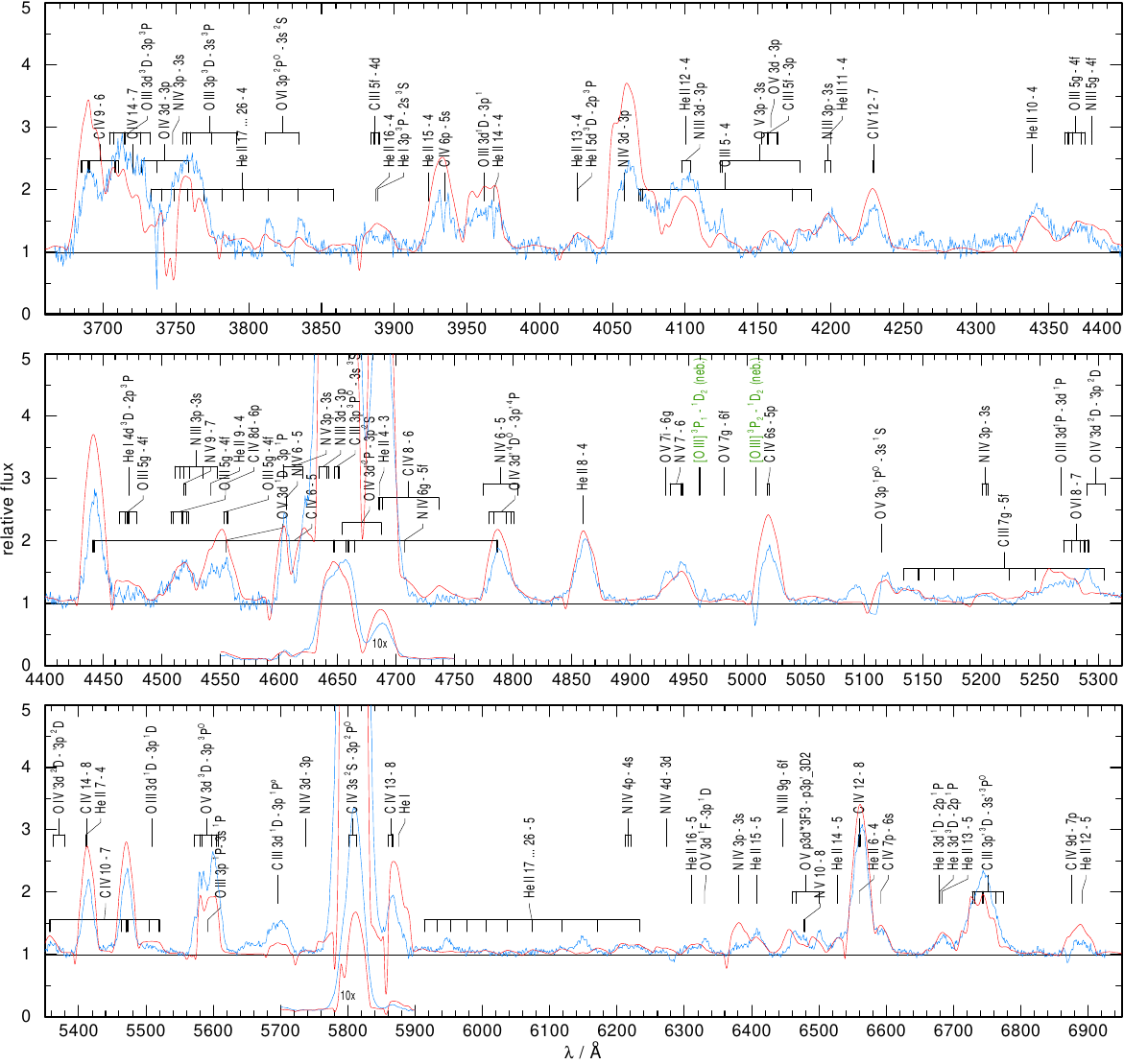}
 \caption{HET/LSR2 spectrum (blue line) of the [WC] central star of RaMul~2. Overplotted (red line) is the best-fitting model, with parameters given in Table\,\ref{tab:results}.}
\label{fig:ramul2}
\end{figure*}

\section{Atmospheric analysis}\label{sect:analysis}

\subsection{The PG1159 central stars of Abell~25 and StDr~138}

The spectrum of the Abell~25 nucleus (Fig.\,\ref{fig:a25}) is similar to that of the prototype of the PG1159 class, PG\,1159$-$035. The latter has an effective temperature of \Teff = 140\,000\,K, a surface gravity of \logg = 7.0, and elemental abundances of He = 0.33, C = 0.50, and O = 0.17 \citep[mass fractions;][]{WernerHeberHunger1991}. Abell~25 exhibits the principal signature of this spectral type, the broad absorption trough at 4630--4720\,\AA, formed by several \ion{C}{iv} lines and \ion{He}{ii} 4686\,\AA. The two central emission spikes \citep[indicating that the star is of spectral subtype ``E'':][]{Werner1992} and the weak emission of \ion{C}{iv} 5801/5812\,\AA\ indicate a relatively high temperature. This is supported by the appearance of \ion{O}{vi} lines, namely the doublet at 3811/34\,\AA\ in absorption and a feature at 5290\,\AA\ with absorption wings and a central emission core, stemming from a few lines between levels with principal quantum numbers $n= 7-8$. The spectrum of StDr~138 (Fig.\,\ref{fig:stdr138}) is rather similar to that of Abell~25. However, the presence of  \ion{O}{vi} is doubtful.\footnote{There are, to our knowledge, no published spectra of the nebulosity surrounding StDr~138. We examined the ``sky'' spectrum from our LRS2 observation. The spectrum is very noisy, given the low surface brightness, small field of view, and the fact that the central star is located in a nebular cavity, but we detect very weak H$\beta$, [\oiii] $\lambda$5007, and H$\alpha$ emission.}

For the spectral analysis of the two PG1159 stars we computed a small grid of non-LTE
model atmospheres of the type introduced by
\citet{werner14}. It was calculated using the T\"ubingen
Model-Atmosphere Package (TMAP) for non-LTE plane-parallel models in
radiative and hydrostatic equilibrium \citep{wernertmap2003}. The
constituents of the models are helium, carbon, and oxygen. The grid covers the
range \Teff = 130\,000--180\,000\,K in steps of 10\,000\,K, and \logg =
6.2--7.0 in steps down to 0.2\,dex. The abundances of the chemical elements of the models range between He = 0.33--0.56, C = 0.36--0.56, and O = 0.02--0.17 in different step sizes down to 0.03.

\begin{figure*}
 \centering  \includegraphics[width=0.7\textwidth]{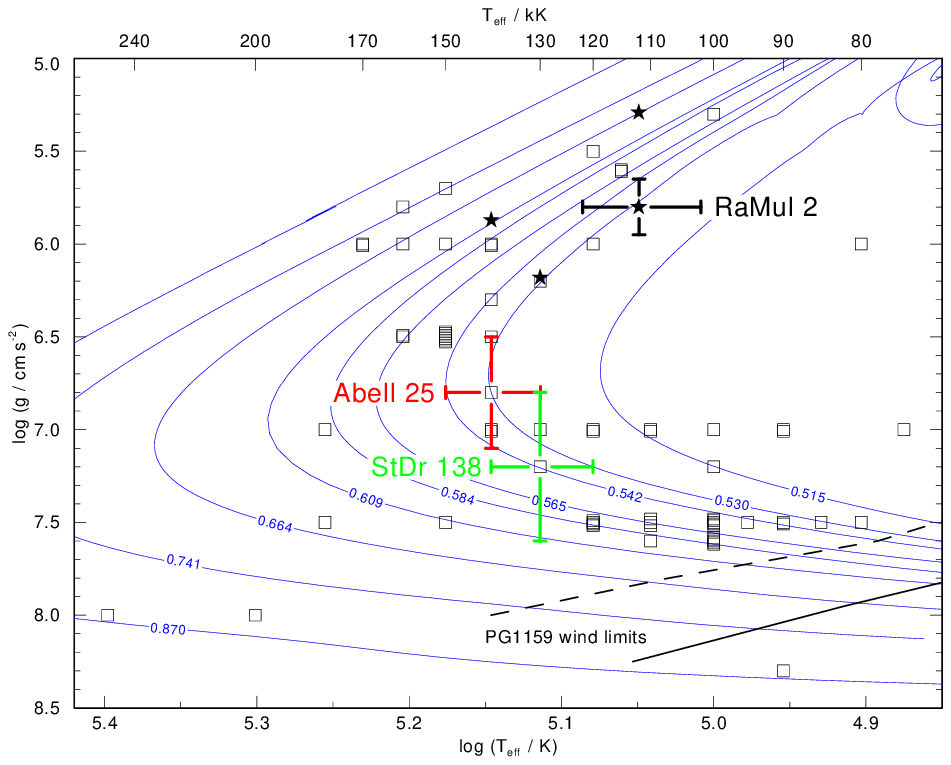}
 \caption{Positions of our program stars in the Kiel diagram, 
      together with all known PG1159 stars (squares; see footnote\,\ref{footpg1159list}) and three [WCE] stars (star symbols; NGC 1501, NGC 2371, and NGC 6905; see Sect.\,\ref{subsect:ramul2}). Blue lines are VLTP post-AGB evolutionary tracks
      by \cite{MillerBertolamiAlthaus2006} labelled with the mass in solar units. 
      The black line indicates the PG1159 wind limit
  from \citet{UnglaubbBues2000}, meaning that the mass-loss
  rate of the radiation driven wind at this position of the evolutionary tracks
  becomes so weak that gravitational settling of heavy elements
  is able to remove them from the atmosphere. Thus, no PG1159
  stars should be found at significantly lower temperatures. The dashed line is the
  wind limit assuming a mass-loss rate that is ten times lower.}
\label{fig:gteff}
\end{figure*}

\begin{figure*}
    \begin{center}
        \includegraphics[width=0.7\textwidth]{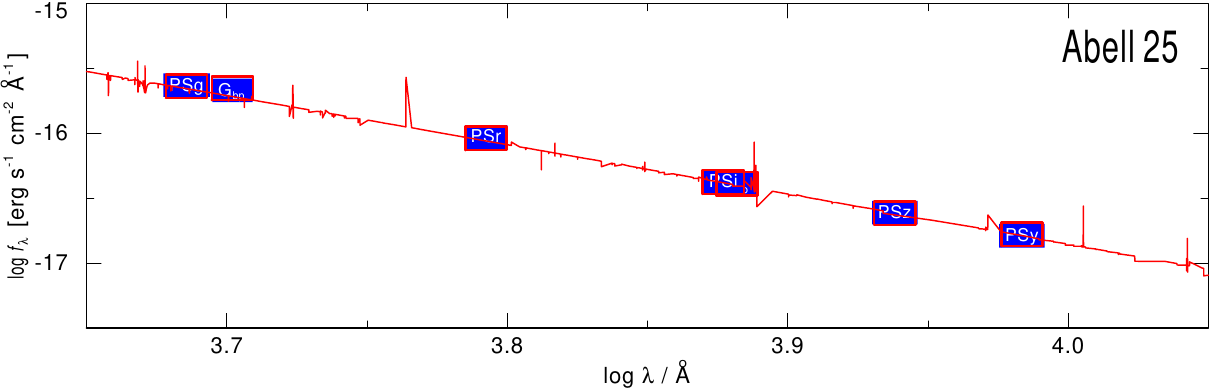}\vspace{1mm}
        \includegraphics[width=0.7\textwidth]{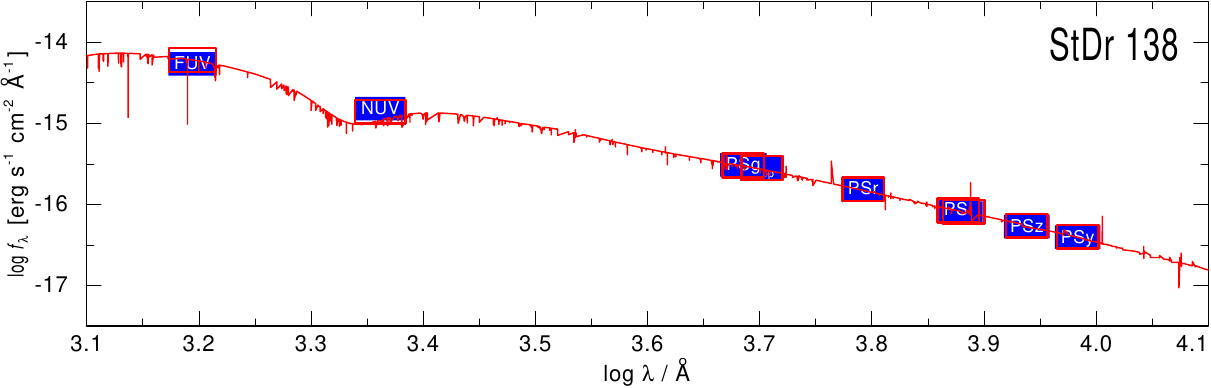}\vspace{1mm}
        \includegraphics[width=0.7\textwidth]{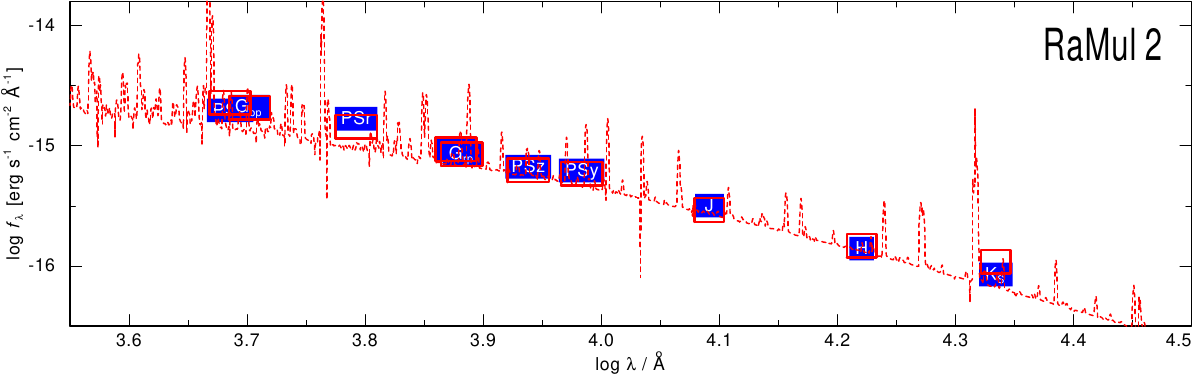}
        \caption{SEDs of our three target stars. Observed photometry is shown by blue boxes and the spectrum of the best fitting model by a red dashed line. Photometry from \textit{GALEX} (FUV and NUV), \textit{Gaia} DR3 (G$_{\rm bp}$), Pan-STARRS (PSg, PSr, PSi, PSz, Psy), and 2MASS (J, H, K$_{\rm S}$). For comparison the resulting photometric values from the models are also shown as red boxes.
        }\label{fig:sed}
    \end{center}
\end{figure*}

The best-fit model for Abell~25 from our grid (as shown by the red line in Fig.\,\ref{fig:a25}) was determined by eye, and it has parameters which are indeed rather similar to those of the PG1159 prototype: \Teff = $140\,000\pm10\,000$\,K, \logg = $6.8\pm0.3$, He = $0.46\pm0.10$, C = $0.46\pm0.10$, and O = $0.08\pm0.05$ (see Table~\ref{tab:results} below). The parameters and their error limits were estimated as follows. Increasing the temperature to 150\,000\,K, the \ion{O}{vi} doublet at 3811/34\,\AA\ becomes too weak and the emission lines of \ion{O}{vi} 5291\,\AA\ and \ion{C}{iv}~5801/5812\,\AA\ become too strong. Decreasing the temperature to 130\,000\,K causes the \ion{O}{vi} feature at 5290\,\AA\ to tend to disappear and the \ion{C}{iv}~5801/5812\,\AA\ emissions become too weak. Additionally, the pair of absorption lines of \ion{He}{ii}~5412\,\AA\ and \ion{C}{iv}~5471\,\AA\ becomes too strong. The surface gravity is constrained, e.g., by the shape of the wings of the lines forming the absorption trough, and of the line pair at 5412\,\AA\ and 5471\,\AA. The He/C abundance ratio is found from the relative strength of the lines in the absorption trough and the 5412\,\AA\ and 5471\,\AA\ line pair. The oxygen abundance is found from the strengths of the \ion{O}{vi} lines at 3811/34\,\AA\ and 5290\,\AA. 

For StDr~138 we found very similar values for temperature and gravity as for Abell~25: \Teff = $130\,000\pm10\,000$\,K, \logg = $7.2\pm0.4$, He = $0.46\pm0.10$, C = $0.46\pm0.10$, and O $\leq 0.08$ (Table~\ref{tab:results}). The model shown in Fig.\,\ref{fig:stdr138} has an oxygen abundance of O = 0.08. The \ion{O}{vi} lines cannot be identified beyond doubt because the S/N of the spectrum is poorer than that from Abell~25. Therefore we adopt this value as an upper abundance limit. 

Some PG1159 stars have nitrogen in oversolar abundance in their atmospheres \citep[of the order N = 0.01; the solar value is 0.00069;][]{Asplund+2009}. This is significant because it signals that the star had experienced a very late thermal pulse (VLTP) in the past, and not a late thermal pulse \citep[LTP; e.g.,][]{werner2006}. In the temperature range relevant for Abell~25 and StDr~138, nitrogen would be detectable through the 4604/20\,\AA\ doublet and a multiplet at 4945\,\AA. These lines are not seen in the observations, and we derive upper limits of N $<$ 0.01 and N $<$ 0.02, respectively. Spectra with better S/N would be necessary to decide whether nitrogen is present in oversolar amounts. 

For the hydrogen abundance we determined an upper limit. The Balmer lines are blended with the \ion{He}{ii} lines of the Pickering series. We find H $\lesssim 0.05$, since for higher H abundances the \ion{He}{ii}~4859\,\AA{}/H$\beta$ blend becomes stronger than observed.

The positions of both central stars in the $\log g$--\Teff diagram are displayed in Fig.\,\ref{fig:gteff}. Stellar radii and masses are determined via interpolation from evolutionary tracks. We employed tracks for post-Asymptotic Giant Branch (AGB) stars that experienced a VLTP \citep{MillerBertolamiAlthaus2006}. Then we calculated the luminosity from the radius and the effective temperature via $L/L_\odot = (R/R_\odot)^2(T_\mathrm{eff}/T_{\mathrm{eff},\odot})^4$. The results are listed in Table\,\ref{tab:results}. The masses of both stars are similar and relatively low, $0.53^{+0.03}_{-0.01}$~\Msol\ (Abell~25) and $0.54^{+0.09}_{-0.02}$~\Msol\ (StDr~138). Both masses are lower than the mean mass of field WDs \citep[$0.61$~\Msol,][]{Kepler2016}.

We derived the interstellar reddening of the central stars of Abell~25 and StDr~138 by fitting the spectral energy distributions (SEDs) of the final model-atmosphere spectra to the observed colors. We use the reddening law of \cite{Cardelli1989} with the color excess $E(B-V)$, apply it to the final model flux and dilute the flux according to the distance $d_{\rm spec}$, such that we get a sufficient agreement with the observed photometry. Fig.\,\ref{fig:sed} compares the reddened theoretical SEDs with the measured photometry. The optimum values for $E(B-V)$ and $d_{\rm spec}$ are inferred via fitting with the reduced $\chi^2$, which also gives us the $1\sigma$ uncertainties. The results are given in Table\,\ref{tab:results}. The spectroscopic distances $d_{\rm spec}$ are found from the relation $f_\lambda = F_\lambda \pi (R/d_{\rm spec})^2$, where $f_\lambda$ is the observed flux distribution and $F_\lambda$ is the (reddened) astrophysical flux from the model atmosphere. For Abell~25 we find $d_{\rm spec}=4592^{+2109}_{-1345}$~pc which is in statistical agreement with the \Gaia\/ parallax distance $d_{\rm Gaia}=2423^{+983}_{-513}$~pc \citep{2021AJ....161..147B}. For StDr~138 we find $d_{\rm spec}=1528^{+1038}_{-595}$~pc which, again, is in statistical agreement with the \Gaia\/ parallax distance $d_{\rm Gaia}=1056^{+237}_{-152}$~pc \citep{2021AJ....161..147B}.


\begin{table*}
\begin{center}
\caption{Parameters of the central stars of Abell~25, StDr~138, and RaMul~2.\tablefootmark{a}}
\label{tab:results}
\begin{tabular}{lccc} 
	\hline\hline
Parameter            & Abell 25                  & StDr 138                  & RaMul 2 \\
	\hline
\noalign{\smallskip}
spectral type        &  PG1159/E                 & PG1159/E                  & [WC4-5] \\
\Teff (K)            & $140\,000\pm10\,000$      & $130\,000\pm10\,000$      & $112\,000\pm10\,000$ \\
\logg (cm s$^{-2}$)  & $6.8\pm0.3$               & $7.2\pm0.4$               & $5.8\pm0.16$\tablefootmark{b}\\ 
\noalign{\smallskip}
H                    & $\leq0.05$                & $\leq0.05$                &$\leq0.03$ \\
He                   & $0.46\pm0.10$             & $0.50\pm0.10$             & $0.40\pm0.10$\\
C                    & $0.46\pm0.10$             & $0.50\pm0.10$             & $0.50\pm0.10$\\
N                    & $\leq 0.01$               & $\leq0.02$                & $0.02_{-0.01}^{+0.03}$\\[0.1cm]
O                    & $0.08\pm0.05$             & $\leq0.08$                & $0.05_{-0.02}^{+0.05}$\\
\noalign{\smallskip}
$M$ (\Msol)          & $0.53^{+0.03}_{-0.01}$    & $0.54^{+0.09}_{-0.02}$    & 0.53\tablefootmark{c}\\     
\noalign{\smallskip}
$R$ (\Rsol)          & $0.048^{+0.022}_{-0.014}$ & $0.031^{+0.021}_{-0.012}$ & $0.149\pm0.026$ \\
\noalign{\smallskip}
$L$ (\Lsol)          & $797^{+1438}_{-500}$      & $247^{+698}_{-180}$       & $3150^{+1200}_{-1020}$ \\
\noalign{\smallskip}
$R_\text{t}$ (\Rsol) &          --               &         --                & $1.45^{+3.55}_{-0.19}$ \\
\noalign{\smallskip}\Mdot (\Msol\ yr$^{-1}$)& -- &         --                &$(4.2\pm1.1)\times10^{-7}$ \\
\noalign{\smallskip}
$\varv_\infty$ (km s$^{-1}$)&        --          &         --                & $1000\pm100$ \\
\noalign{\smallskip}
$d_{\rm spec}$ (pc) & $4592^{+2109}_{-1345}$     & $1528^{+1038}_{-595}$     & $8850^{+530}_{-780}\,$\tablefootmark{d} \\
\noalign{\smallskip}
$d_{\rm Gaia}$ (pc) & $2423^{+983}_{-513}$       & $1056^{+237}_{-152}$      & $6386^{+1073}_{-996}$\\
\noalign{\smallskip}
$E(B-V)$ (mag)      & $0.05^{+0.03}_{-0.02}$     &  $0.33^{+0.03}_{-0.02}$   & $0.46^{+0.10}_{-0.12}$\\
\noalign{\smallskip}
  \hline
\end{tabular}
\tablefoot{  
\tablefoottext{a}{Element abundances given in mass
    fractions. Stellar masses of the PG1159 stars derived from VLTP tracks
    (Fig.\,\ref{fig:gteff}). \Teff and $R$ for RaMul~2 defined at $\tau_{\rm Ross}=20$. The {\it Gaia} distances were taken from
    \cite{2021AJ....161..147B}. } 
\tablefoottext{b}{Computed from $R$ and $M$.}    
\tablefoottext{c}{From Fig.\,\ref{fig:ramul2_hrd}, adopted value for the \textit{PoWR} model is 0.6\,\Msol.}    
\tablefoottext{d}{Computed from an assumed value for the luminosity of $L=6000\,\Lsol$.}
    } 
\end{center}
\end{table*}

\subsection{The [WC] central star of RaMul~2}

\subsubsection{Spectral classification}

The emission-line spectrum of the central star of RaMul~2, shown in Fig.~\ref{fig:ramul2}, is that of a [WC] star, i.e., a WR star dominated by carbon features. In this notation, the square brackets indicate that it is a low-mass Population~II star, as distinguished from the massive Population~I WR stars that show very similar spectra \citep{Vanderhucht1981}. WR-type central stars cover the spectral subtypes [WO1]--[WO4], which are the hottest objects \citep[\Teff $> 100\,000$\,K; e.g.,][]{Rubio2022}, and [WC4]--[WC12], with the latest subtypes having  temperatures down to 20\,000\,K \citep{Leuenhagen1996}. It is thought that the WR PN nuclei form an evolutionary post-AGB sequence, starting from the coolest and most luminous objects to the fainter and hottest ones \citep{KoesterkeHamann1997}, which subsequently transform into PG1159 stars \citep{Mendez1991,WernerHeberHunger1991,WernerHeber1991}.

\begin{table}
\centering
\caption{Equivalent widths of spectral lines of the RaMul~2 central star used for classification.}
\label{tab:ramul2_class}
\begin{tabular}{llcc} 
        \hline\hline 
Ion &         $\lambda$    &       $-W_\lambda$ & \multicolumn{1}{c}{ratio \ion{C}{iv}5801/12} \\
    &                [\AA] & [\AA]          &  \multicolumn{1}{c}{$=100$}\\
        \hline
\ion{C}{ii}   & 4267    & 0\tablefootmark{a}    & 0 \\
              & 6461    & 0\tablefootmark{a}    & 0 \\
\ion{C}{iii}  & 4649    & 428  & 48 \\
              & 5696    &  16  & 1.8\\
              & 6730    &  49  & 6.4\\
\ion{C}{iv}   & 5801/12 & 900  & 100\\
              & 5470    & 25   & 2.8\\
\ion{O}{iii}  & 5592    & 50  &  5.6         \\
\ion{O}{vi}   & 3811/34 & 9   & 1 \\
              & 5290    & 6   & 0.7\\
        \hline
\end{tabular}
\tablefoot{
\tablefoottext{a}{absent}
}
\end{table}

It was realised long ago that there is a conspicuous gap in the [WC] subtype sequence lying between the late and early subtypes ([WCL] and [WCE]), with very few objects found among the intermediate subtypes in the range [WC5]--[WC7] \citep{Mendez1982}. A similar phenomenon is not shown by the massive WR star counterparts. See, for example, the central-star catalogue of \cite{Weidmann2020}, their Fig.\,6. However, there is not a counterpart of the gap in the \Teff-distribution of [WC] stars, suggesting that specific atmospheric conditions are necessary for [WC] central stars to exhibit a [WC5]--[WC7] spectrum \citep{Weidmann2020}. Out of about one hundred classified Galactic [WC] central stars, there is not a single one with class [WC6] or [WC7], and only two with class [WC5]. Five objects were classified as [WC5-6]. 

We give measurements of equivalent widths of features useful for spectral classification of RaMul~2 in Table~\ref{tab:ramul2_class}. \citet{Crowther1998} list quantitative spectral-classification criteria for the subtypes of WC stars. For subtype WC4 the primary criterion is a logarithmic equivalent-width ($\log W_\lambda$) ratio of \ion{C}{iv}~5808\,\AA{}/\ion{C}{iii}~5696\,\AA{} $\ge 1.5$, and a secondary criterion is a $\log W_\lambda$ ratio of  \ion{C}{iii}~5696\,\AA{}/\ion{O}{v}~5590\,\AA\, $\leq -0.4$. Our measurements of these two ratios for RaMul~2 are 1.7 and $-0.5$, respectively, consistent with a spectral type of [WC4]. We also note the similarity of its spectrum to that of the central star of the PN NGC\,5315, classified [WC4] by \cite{Tylenda1993}.

However, according to an updated classification scheme for [WC] stars introduced by \citet{Acker2003}, RaMul~2 has line strengths of \ion{C}{iii}~5696\,\AA{} and \ion{C}{iii}~6730\,\AA{} relative to \ion{C}{iv}~5801/12\,\AA{} (see Table~\ref{tab:ramul2_class}) that indicate a later spectral subtype of [WC5-6], while other criteria, including the absence of \ion{C}{ii} lines, support a classification of [WC4]. Finally, \citet{Mendez1982} give as a criterion for [WC5] vs.\ [WC4] that \ion{C}{iii}~5696\,\AA{} is stronger than \ion{O}{vi}~5290\,\AA{}, as it is for RaMul~2. They also list as a criterion the strength of the \ion{O}{v}~5595\,\AA{} line, but for RaMul~2 this feature is actually a blend of \ion{O}{v} and \ion{O}{iii} and therefore ambiguous. 
Moreover, its spectrum is very similar to that of the central star of SMP~61 in the Large Magellanic Cloud (LMC), classified as [WC5-6] by \citet{Stasinska2004}.

The spectrum of RaMul~2 also shows strong nitrogen emission lines, which are typical of WN-type or WC/WN-type stars \citep[cf.][]{Vanderhucht1981,Hamann2006, Sander2012}. When applying a WN-type classification scheme \citep[e.g.,][]{Smith1968} for massive stars, one would assign a [WN5] subtype to RaMul~2, because of the apparent almost equal line strengths from \ion{N}{iii}, \ion{N}{iv}, and \ion{N}{v}. However, all of the nitrogen lines in the spectrum of RaMul~2 are heavily blended with other metal lines, in contrast to the spectra of massive WN stars or the low-mass WN-type stars, e.g., IC~4663 \citep{Miszalski2012}, Abell~48 \citep{Todt2013}, and PB~8 \citep{Todt2010}, therefore masking the real strengths of the N lines. Moreover, WN-type stars usually have much weaker carbon and oxygen lines, as the chemical abundances of carbon and oxygen in such stars are much lower than in WC-type stars. 
Hence, we refrain from applying a WN classification scheme to RaMul~2.

Our final conclusion is that we assign a spectral type of [WC4-5] to RaMul~2, meaning that it is located right at the hot edge of the gap in the spectral-type distribution of [WC] stars.
As mentioned above, RaMul~2 was previously classified by \cite{Acker2015} as [WO4] according to the scheme of \cite{Acker2003}.

\subsubsection{Modeling the spectrum}

The strong and broad emission features in the spectrum of the central star of RaMul~2 indicate a strong stellar wind. To analyse its spectrum, we use the most recent version\footnote{2023 August 17} of the non-LTE Potsdam stellar atmosphere code \citep[\textit{PoWR};][]{Grafener2002,Hamann2004}\footnote{\url{https://www.astro.physik.uni-potsdam.de/PoWR}} for expanding atmospheres. Details on the code's performance are given by \citet{Todt2015}, and applications to WR-type central stars can be found in the Potsdam group's recent papers \citep[see][]{Toala2019,GG2020,GG2022}.

The basic assumptions of the \textit{PoWR} code are spherical symmetry and stationarity of the radial outflow, where we have to give two of the three quantities $L$, $T_{\rm eff}$, and $R$ for the inner boundary.
The radiative-transfer equation is solved in the comoving frame of the expanding atmosphere, iteratively with the equations of statistical equilibrium  and radiative equilibrium. In the subsonic part,  we assume a velocity field corresponding to a quasi-hydrostatic density stratification according to the continuity equation. For the supersonic part of the wind, we prescribe the velocity field $\varv(r)$ by a so-called $\beta$-law, where the free parameter $\beta$ for WR stars is usually set to $\beta=1$. 
From the widths of the emission lines, specifically from the \ion{He}{ii}~6560\,\AA{} line, we infer a terminal velocity of the stellar wind of $\varv_\infty = 1000\pm100$\,km\,s$^{-1}$. Additional line broadening by microturbulence, with $\varv_\text{turb}=100$\,km\,s$^{-1}$, is included in our models. We adopt a stellar mass of $M=0.6\,\Msol$, which is close to the mean mass of WDs \citep[$0.61$~\Msol,][]{Kepler2016}. Usually, the value of $M$ has no noticeable influence on the synthetic wind spectra.

The emission-line spectra of WR stars are mainly generated through recombination processes in their dense stellar winds. Therefore, the continuum-normalized spectrum exhibits a valuable scale-invariance. Specifically, for a given effective temperature\footnote{In this paper we use the values of \Teff\ and stellar radius $R$ defined at $\tau_{\rm Ross}=20$ in the \textit{PoWR} models.} and chemical composition, the equivalent widths of the emission lines are primarily determined by the ratio of the volume emission measure of the wind to the surface area of the star in a first approximation.
An equivalent quantity, which has been introduced by 
\cite{schmutz1989}, is the so-called transformed radius 
\begin{equation}
 R_\text{t} =
 R\left[\left. \frac{\varv_\infty}{2500\,\text{km}\,\text{s}^{-1}}
 \right/ \frac{\dot{M}\sqrt{D}}{10^{-4}\,\text{M}_\odot
 \,\text{yr}^{-1}} \right]^{2/3}~~.  
 \label{eq:pb8-transradius}
\end{equation}

Various combinations of stellar radii $R$ and mass-loss rates $\dot{M}$ can result in equivalent emission-line strengths. The presented formulation of invariance extends to include the micro-clumping parameter $D$, representing the density contrast between wind clumps and a continuous wind with the same mass-loss rate. As a result, empirically derived mass-loss rates from fitting the emission-line spectrum are contingent on the chosen value of $D$. This parameter can be constrained by fitting the extended electron-scattering wings observed in strong emission lines
\citep[e.g.,][]{Hamann1998}. We use a value of $D=10$ \citep[cf.][]{Todt2008}, as models with a homogeneous wind give too strong electron-scattering wings. 

The spectrum of RaMul~2 (Fig.~\ref{fig:ramul2}) shows emission lines of different ionization stages of carbon (\ion{C}{iii}, \ion{C}{iv}), helium (\ion{He}{i}, \ion{He}{ii}), oxygen (\ion{O}{iii--vi}), and nitrogen (\ion{N}{iii--v}), where the relative line strengths are sensitive to \Teff\ and wind density. Our best-fitting model is a compromise model which reproduces qualitatively all spectral lines. Models with a larger value for $R_\text{t} = 5.0\,$\Rsol{} (i.e., lower wind density) and \Teff $\approx$ 100\,000\,K result in a better fit to the \ion{He}{i} lines and the oxygen lines, specifically to the \ion{O}{vi}~5290\,\AA{} line, while models with a higher temperature (\Teff $\approx$ 125\,000\,K) and a lower $R_\text{t}=1.26\,$\Rsol\ yield a better agreement for the \ion{C}{iv} lines and the \ion{O}{vi}~3811/3834\,\AA\ doublet. However, these models have a thick wind, so within the $T_{\rm eff}$--$R_\text{t}$ diagram they lie in the regime of parameter degeneracy, where the spectra depend only on the product $R_\text{t} T_{\rm eff}^2$ \citep{Hamann2003} and therefore also models with even larger values of \Teff might still give a qualitatively sufficient fit. 

When comparing the spectrum and also the inferred values of RaMul~2 to those of the massive WC stars from \citet{Sander2012}, we first notice that almost all objects with spectral subtype WC5 and WC6 in their paper were fitted with the same \textit{PoWR} model, which has \Teff $\approx 79\,$kK and $\log(R_\text{t}/\Rsol) = 0.5$, indicating that also the spectra look very similar. Moreover, the objects WR\,11 (WC5) and WR\,132 (WC6), as well as the objects of spectral subtype WC4, were fitted with \textit{PoWR} models which lie in the $T_{\rm eff}$--$R_\text{t}$ diagram on the same diagonal as the model for the other WC5 and WC6 stars. Consequently, their spectra also look very similar. Specifically, the WC4 stars WR\,144 and WR\,52 were fitted with a \textit{PoWR} model that has \Teff $\approx 112\,$kK and $R_\text{t}\approx 1.59\,$\Rsol, thus very close to the values we derived for RaMul~2. However, in contrast to the results of the spectral analysis of the massive WC stars, we find a lower value for $\varv_\infty$ (1000\,km/s vs.\ $\approx 2000$\,km/s) and also obtain a slightly better fit to the optical spectrum, specifically for the \ion{C}{iii}/\ion{C}{iv} blend at $\approx4650\,\AA$.

Our atmosphere models include model atoms of helium, carbon, oxygen, nitrogen, neon, hydrogen, and the iron-group elements (Sc--Ni). Similar to the analysis of the two PG1159 stars described above, the chemical abundances of helium and carbon are derived with help of the diagnostic line pair \ion{He}{ii} 5412\,\AA\ and \ion{C}{iv} 5471\,\AA. 
From their relative strengths (ratio of the equivalent widths) we infer He = 0.4 and C = 0.5. From the absence of strong neon emission lines we can only infer an upper limit of Ne $<$ 0.03, because models with higher neon abundances exhibit the \ion{Ne}{iv} 3821\,\AA\ line, which is not observed.

To reproduce the multiplet blend of \ion{O}{iii} and \ion{O}{v} at about 5600\,\AA\ and the \ion{O}{iii} lines at $\approx$3750--3850\,\AA{}, an oxygen abundance of O = 0.10 is required. However, other oxygen lines, like \ion{O}{iii} 3962\,\AA{}, \ion{O}{iii} 5268\,\AA{}, and \ion{O}{v} 6460, 6466, 6500\,\AA{} then become much too strong in the models as compared to observations. These lines are better reproduced with O = 0.03. The best overall fit is achieved with O = 0.05.
The spectrum of RaMul~2 shows a number of strong nitrogen lines, indicating that this species might be enriched. In fact, our best fit is obtained by models with N = 0.02, which especially fits well the \ion{N}{iv} multiplet at 6212,  6215, and 6220\,\AA{}. The multiplets of \ion{N}{v} at 4934, 4943, 4944, 4945 and at 4604, 4620\,\AA{} are best reproduced with N = 0.05, while, e.g., the emission lines of \ion{N}{iii} 6381\,\AA{} and \ion{N}{iv} 4058\,\AA{} require a lower abundance of N = 0.01. The best overall fit to the nitrogen lines is achieved with N = 0.02. 

We also attempted to determine the hydrogen abundance, but could only derive an approximate upper limit because, as in the case of the PG1159 stars, the Balmer lines are blended with the \ion{He}{ii} lines of the Pickering series. Moreover, almost all of the Pickering \ion{He}{ii} lines are blended with metal lines, so the Pickering decrement cannot be measured properly to estimate the hydrogen abundance. We find that H $\lesssim 0.03$, as for higher H abundances the \ion{He}{ii}~4859\,\AA{}/H$\beta$ and \ion{He}{ii}~6560\,\AA{}/H$\alpha$ blends become stronger than observed.
An ultraviolet spectrum would be necessary to determine the iron abundance. For our models we adopted solar abundances for the iron-group elements, in particular Fe = 0.0014 \citep{Asplund+2009}. 

Finally, the model spectrum with absolute fluxes is subjected to interstellar extinction, using the reddening law of \cite{Cardelli1989}, and diluted according to the \textit{Gaia} parallax distance from \citet[][see Table\,\ref{tab:results}]{2021AJ....161..147B}. The values for $E(B-V)$ and the stellar luminosity $L$ are inferred by fitting the synthetic spectrum to the observed photometry, where we make use of the scale-invariance of WR spectra (see above). The optimum values are obtained for a reduced $\chi^2_\text{red}$ of 1, where the values of $E(B-V)$ and $L$ with $\chi^2_\text{red}=2$ give the $1\sigma$ uncertainty interval (Table\,\ref{tab:results}). The SED of RaMul~2 is shown in the bottom panel of Fig.\,\ref{fig:sed}.

WR spectra normally do not have photospheric absorption lines, which could otherwise allow the determination of $\log g$ and hence of the stellar radius $R$ for a given mass, as we did for the Abell~25 and StDr~138 central stars. Consequently, for the [WC] central star of RaMul~2 the stellar luminosity $L$ must be either assumed or can be obtained from the given distance of the object. If we assume a typical luminosity for a central star of $L=6000\,\Lsol$ \citep[e.g.,][]{schoenberner2005,miller-bertolami2007}, we can also calculate a spectroscopic distance for RaMul~2 of about 8850\,pc.

The position of the central star of RaMul~2 in the Kiel diagram is shown in Fig.\,\ref{fig:gteff}. We note that there the position of RaMul~2 compared to the tracks is inconsistent with the mass of $0.6\,\Msol$, which we assumed for our \textit{PoWR} models, but as stated before, the value of $\log g$ cannot be inferred independently, but one must assume a stellar mass. This situation might improve with the hydrodynamically consistent \textit{PoWR} models \citep[e.g.,][]{graefener2005,graefener2008,Sander2017}, where the mass-loss rate and the terminal velocity are predicted from the input values of $L$ and $M$. For a better comparison we show the positions of RaMul~2 and the other two objects in the Hertzsprung-Russell diagram (HRD) together with theoretical tracks from \citet{MillerBertolamiAlthaus2006} in Fig.\,\ref{fig:ramul2_hrd}. Here, RaMul~2 lies on the track for a remnant mass of $0.530\,$\Msol. That is consistent with its position in the Kiel diagram. 

Moreover, [WC] stars and PG1159 stars populate the same region of the Kiel diagram, where [WC] stars have strong winds and PG1159 stars show no signs of a stellar wind. This is likely due to higher values of $L/M$ for [WC] stars, putting them closer to the Eddington-limit (as for massive stars), and maybe also caused by larger wind-driving opacities due to lower temperatures. However, there are also [WC] stars with temperatures similar to those of the PG1159 stars. Hence, [WC] winds might form only in a specific parameter range of $L,M,T_{\rm eff}$ and for metal-rich abundances. That needs to be explored with help of hydrodynamically consistent model atmospheres. 

\begin{figure}
    \centering
    \includegraphics[width=\columnwidth]{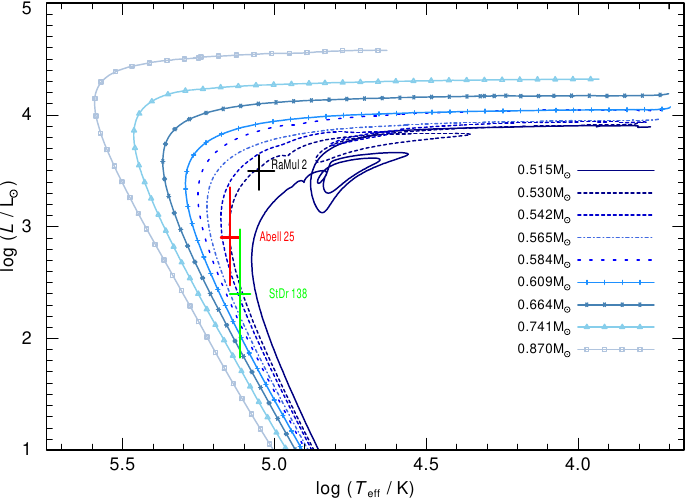}
    \caption{HRD with the post-VLTP tracks from \citet{MillerBertolamiAlthaus2006} for different remnant masses. The positions of our three central stars are shown with their error bars.}
    \label{fig:ramul2_hrd}
\end{figure}


\section{Summary and conclusions}\label{sect:summary}

\subsection{The PG1159 central stars of Abell~25 and StDr~138}

We found that the central stars of Abell~25 and StDr~138 are two new PG1159 stars. They are of relatively low mass (0.53 and 0.54\,\Msol), and with \Teff = 140\,000 and 130\,000\,K they have just passed the maximum effective temperature of their post-AGB evolution (Figs.\,\ref{fig:gteff} and~\ref{fig:ramul2_hrd}). They join the PG1159 class, which now comprises 71 objects.\footnote{According to an unpublished list based on \cite{werner2006} and maintained by the first author.\label{footpg1159list}} Twenty-seven of them have an associated PN\null. The other PG1159 stars likely also ejected PNe, but they have now dispersed (but see below).

Generally, the hydrogen--deficiency in PG1159 stars is thought to be caused by a (V)LTP \citep{Iben1983}, an event originally predicted in evolutionary calculations by \citet{Fujimoto1977} and \citet{Schoenberner1979}. We can distinguish between three scenarios, namely, an LTP or a VLTP, depending on whether the pulse occurs during the pre-WD post-AGB evolution or during the later WD-cooling phase, or an AGB final thermal pulse (AFTP) occurring immediately before the departure of the star from the AGB \citep[e.g.,][]{werner2006}. The expected abundances of H, He, C, N, and O are different for the three cases. An AFTP causes the dilution of hydrogen by envelope convection to an extent that hydrogen can still be detected spectroscopically, namely, of the order H = 0.2 \citep[e.g.,][]{Loebling2019}. 
These are the so-called hybrid-PG1159 stars \citep{Napiwotzki1991} and four such objects are known \citep[two of them have a PN, namely, Abell~43 and NGC~7094;][]{Loebling2019}.
In the LTP and VLTP cases, on the other hand, hydrogen disappears almost completely. The difference between these two events is that hydrogen is highly diluted in the LTP, but in the VLTP it is ingested and burned to helium. Observationally, the results of these two events can be distinguished, because in the VLTP case H-burning will increase the nitrogen abundance to as high as a few percent. Indeed, there is a dichotomy of PG1159 stars showing such high amounts of nitrogen while others do not. Nine PG1159 stars with high N abundance are known \citep[see compilation in Table\,2 of][]{Sowicka2023}, but we note that good spectra are needed to discover nitrogen lines. Unfortunately, the upper limit for the N abundance in our two new PG1159 stars is N = 0.02 and 0.01, so that we are unable to decide whether they experienced a LTP or a VLTP\null. 
To settle this question, spectra with higher S/N are required because the nitrogen lines are rather weak.

We add that recently two other pathways have been identified that can also lead to objects of spectral type PG1159. They mark the drastic end of two WDs in a close-binary system. 
First, two hot helium-rich subdwarfs with considerable amounts of carbon and oxygen were discovered \citep{WernerReindl2022}. They are probably the outcome of a binary WD merger resulting in a hot subdwarf with a surface composition which is similar to the outcome of a VLTP event \citep{MillerBertolami2022}.
Second, a recently detected hypervelocity star exhibits a PG1159-type spectrum. The object was identified as a runaway WD companion from a Type Ia supernova \citep{Elbadry2023}. It has a carbon-oxygen dominated atmosphere \citep{WernerElbadry2024} composed of debris material from the exploded CO WD primary. Both of these two alternative scenarios contribute, perhaps in a small fraction, to the PG1159 population which are not associated with a PN.

The PG1159 central stars of Abell~25 and StDr~138 are located within the GW~Vir instability strip, close to the blue edge \citep[see Fig.\,1 in][]{Corsico2021}. The prototype of the PG1159 stars (PG\,1159$-$035 = GW~Vir) is also the prototype of the GW~Vir variables. 
It was shown that from a total of 67 PG1159 stars 24 are confirmed low-amplitude non-radial $g$-mode pulsators \citep{Sowicka2023}. From the temperatures and luminosities one could expect photometric variability of our two new PG1159 stars with periods in the range of approximately 5--50 minutes and amplitudes of a few millimag up to about 0.1~mag. However, the GW~Vir instability strip is not pure in the sense that the majority of PG1159 stars located within the strip are non-pulsators. Interestingly it was found from a sample of nine PG1159 stars that only the pulsators have high nitrogen abundances, leading to speculations that pulsators and non-pulsators have different evolutionary histories, namely, VLTP and LTP, respectively \citep{Dreizler1998}. This correlation between nitrogen-overabundance and occurrence of pulsations was investigated recently with a larger sample of PG1159 stars \citep{Sowicka2023}. Five out of 14 pulsators with published N abundances are N-poor, thus, the previously claimed dichotomy has apparently disappeared.

\subsection{The Wolf-Rayet central star of RaMul~2}\label{subsect:ramul2}

The central star of RaMul~2 is of spectral type [WC4-5]. Hence it is located at the limit to the region where intermediate spectral types are very rare. Its spectrum is very similar to the central star of SMP~61 in the LMC \citep{Stasinska2004}, which was classified as [WC5-6]. As is the case for PG1159 stars, some [WC] stars show an enhanced nitrogen abundance, while others do not \citep[e.g.,][]{TodtWDworkshop2015}. A high nitrogen abundance in [WC] stars is therefore also interpreted as the result of a VLTP, and their successors are the N-rich PG1159 stars \citep{WernerHeber1991}. Obviously, with N = $0.02_{-0.01}^{+0.03}$, this is the case for the central star of RaMul~2. SMP~61 on the other hand does not show nitrogen lines (N $< 5\times10^{-5}$), indicating that it experienced a LTP.

Concerning the origin of [WC] stars, it was claimed that there is observational evidence that not all objects can be the result of the final helium-flash events in single stars \citep[see, e.g.,][]{Gorny2008,DeMarco2008}. 
In particular it was questioned whether there is an evolutionary link between the [WCL] and [WCE] subtypes \citep{Gorny2008}.
Inspiral of a low-mass companion, brown dwarf, or a planet into an AGB star was proposed as an alternative \citep{DeMarcoSoker2002,DeMarco2008} to initiate a thermal pulse creating hydrogen--deficiency. 

Furthermore, it was argued that the discovery of apparently frequent [WC] central stars in metal-poor environments \citep[the Galactic bulge and, suggestively, the Sgr dwarf spheroidal Galaxy,][respectively]{Gorny2004,Zijlstra2006} points at the possibility that metallicity could determine the occurrence of [WC] stars.
But we note this discovery could be prone to a selection effect since [WC] central stars are easier to identify because of their bright emission lines. Metallicity was also suspected to be the cause of the observation that Magellanic Cloud [WC] central stars are predominantly of intermediate subtype \citep[e.g.,][]{Monk1988}, while such objects are rare in the Galaxy \citep{Stasinska2004}. The origin of this behaviour could be different wind properties because the wind structure depends on metallicity \citep{Crowther2008}.

Like the two new PG1159 stars we found, the central star of RaMul~2 is also located within the GW~Vir instability strip. It is in the vicinity of the three known [WCE] pulsators. These are the central stars of NGC~1501, NGC~2371, and NGC~6905 \citep{CiardulloBond1996}, and their positions in the Kiel diagram are shown in Fig.\,\ref{fig:gteff}. Their effective temperatures are 112\,000, 130\,000, and 139\,000\,K, respectively \citep{Rubio2022,GG2020,GG2022}, and their variability with periods between about 700 and 5100~\,s was studied in detail by \cite{Corsico2021}. Photometric observations might reveal such pulsations in RaMul~2.

Taken together, our three newly identified hydrogen-deficient stars have, coincidentally, similar masses of about 0.53\,\Msol\ and, hence, represent three different stages of post-AGB evolution of an initially low-mass star. It was shown that Population~II main-sequence stars with initial mass 0.8\,\Msol, i.e., spectral type early-K, evolve to WDs with $M=0.53$\,\Msol\ \citep{Kalirai2009}. Our central stars cover the phase during which the stellar luminosity drops from about 3000 to 250\,\Lsol. This and the contraction of the star within this time interval together with the increase of the surface gravity from \logg = 5.8 to 7.2 causes a throttling of the wind mass-loss rate from $\log \dot{M}/(\Msol\,{\rm yr}^{-1}) = -6.4$ (the value we found for the [WC] nucleus of RaMul~2) to a value that is undetectable in the optical spectra of the PG1159 stars. The two PG1159 stars represent the future of the central star of RaMul~2. Its effective temperature will rise from 112\,000\,K to a maximum value of 140\,000\,K and finally it will enter the WD cooling sequence. Consistently with this picture of the evolutionary sequence of our three targets, the PN around RaMul~2 is physically the smallest of the three (see Sect.~\ref{subsec:targetRaMul2 }), and the StDr~138 PN is the largest (Sect.~\ref{subsec:targetStDr138}).

\begin{acknowledgements} 

We thank the referee for their comments, which helped to improve the paper.

This research has made use of NASA's Astrophysics
Data System and the SIMBAD database, operated at CDS, Strasbourg,
France. This research has made use of the VizieR catalogue access
tool, CDS, Strasbourg, France. This work has made use of data from the
European Space Agency (ESA) mission {\it Gaia}.

We thank the HET queue schedulers and nighttime observers at McDonald Observatory for obtaining the data discussed here.


The Low-Resolution Spectrograph 2 (LRS2) was developed and funded by the University of Texas at Austin McDonald Observatory and Department of Astronomy, and by Pennsylvania State University. We thank the Leibniz-Institut f\"ur Astrophysik Potsdam (AIP) and the Institut f\"ur Astrophysik G\"ottingen (IAG) for their contributions to the construction of the integral-field units.

We acknowledge the Texas Advanced Computing Center (TACC) at The University of Texas at Austin for providing high-performance computing, visualisation, and storage resources that have contributed to the results reported within this paper.

This work has made use of data from the European Space Agency (ESA) mission
{\it Gaia\/} (\url{https://www.cosmos.esa.int/gaia}), processed by the {\it Gaia\/}
Data Processing and Analysis Consortium (DPAC,
\url{https://www.cosmos.esa.int/web/gaia/dpac/consortium}). Funding for the DPAC
has been provided by national institutions, in particular the institutions
participating in the {\it Gaia\/} Multilateral Agreement.

This research has made use of the SIMBAD database, operated at CDS, Strasbourg, France.

\end{acknowledgements}

\bibliographystyle{aa}
\bibliography{aa}

\end{document}